\documentclass[11pt]{article}
\usepackage{amssymb}

\usepackage{epsfig,latexsym}

\def\theequation{\arabic{section}.\arabic{equation}}

\renewcommand{\theequation}{\thesection.\arabic{equation}}

\global\arraycolsep=1pt
\input epsf
\input{tcilatex}
\begin{document}

\font\cmss=cmss10 \font\cmsss=cmss10 at 7pt \hfill \hfill IFUP-TH/2002-38


\vspace{10pt}

\begin{center}
\vskip .5truecm{\Large \textbf{\vspace{10pt}}}

{\Large \textbf{INEQUALITIES FOR TRACE ANOMALIES, LENGTH\ OF\ THE\ RG\ FLOW,
DISTANCE\ BETWEEN\ THE\ FIXED\ POINTS AND IRREVERSIBILITY}}

\bigskip \bigskip \vskip .5truecm

\textsl{Damiano Anselmi}

\textit{Dipartimento di Fisica E. Fermi, Universit\`{a} di Pisa, and INFN}
\end{center}

\vskip 2truecm

\begin{center}
\textbf{Abstract}
\end{center}

I\ discuss several issues about the irreversibility of the RG flow and the
trace anomalies $c$, $a$ and $a^{\prime }$. First I argue that in quantum
field theory: $i$)\ the scheme-invariant area $\Delta a^{\prime }$ of the
graph of the effective beta function between the fixed points defines the
length of the RG flow; $ii$) the minimum of $\Delta a^{\prime }$ in the
space of flows connecting the same UV and IR fixed points defines the
(oriented) distance between the fixed points; $iii$) in even dimensions, the
distance between the fixed points is equal to $\Delta a=a_{\mathrm{UV}}-a_{%
\mathrm{IR}}$. In even dimensions, these statements imply the inequalities $%
0\leq \Delta a\leq \Delta a^{\prime }$ and therefore the irreversibility of
the RG flow. Another consequence is the inequality $a\leq c$ for free
scalars and fermions (but not vectors), which can be checked explicitly.
Secondly, I elaborate a more general axiomatic set-up where irreversibility
is defined as the statement that there exist no pairs of non-trivial flows
connecting interchanged UV and IR fixed points. The axioms, based on the
notions of length of the flow, oriented distance between the fixed points
and certain ``oriented-triangle inequalities'', imply the irreversibility of
the RG flow without a global $a$ function. I conjecture that the RG\ flow is
irreversible also in odd dimensions (without a global $a$ function). In
support of this, I check the axioms of irreversibility in a class of $d=3$
theories where the RG flow is integrable at each order of the large N
expansion.

\vskip 2truecm

\vfill\eject

\section{Introduction}

\setcounter{equation}{0}

At the critical points of the RG flow, the trace anomaly in external gravity
in even dimensions $d=2n$ contains three types of terms, constructed with
the curvature tensors and their covariant derivatives: \textit{i}) terms $%
\mathcal{W}_{i}$, $i=0,1,\ldots ,I$, such that $\sqrt{g}\mathcal{W}_{i}$ are
conformally invariant; \textit{ii}) the Euler density 
\[
\mathrm{G}_{d}=(-1)^{n}\varepsilon _{\mu _{1}\nu _{1}\cdots \mu _{n}\nu
_{n}}\varepsilon ^{\alpha _{1}\beta _{1}\cdots \alpha _{n}\beta
_{n}}\prod_{i=1}^{n}R_{\alpha _{i}\beta _{i}}^{\mu _{i}\nu _{i}}~; 
\]
\textit{iii}) covariant total derivatives $\mathcal{D}_{j}$, $j=0,1,\ldots
,J $, having the form $\nabla _{\alpha }J^{\alpha }$, $J^{\alpha }$ denoting
a covariant current.

The coefficients multiplying these terms are called ``central charges'' and
can be conveniently normalized by the formula 
\begin{eqnarray}
{\Theta _{d\mathrm{=}2n}^{*}} &=&{{\frac{{n}!}{(4\pi )^{n}\,(d+1)!}}\left[ {%
\frac{c_{d}\left( d-2\right) }{4(d-3)}}\mathcal{W}_{0}-{\frac{2^{1-n}}{\,d}}%
a_{d}\mathrm{G}_{d}+{\frac{a_{d}^{\prime }}{(d-1)}}\Box ^{n-1}R+\right. } 
\nonumber \\
&&\qquad \qquad \qquad \qquad \qquad \qquad \qquad \qquad \left.
+\sum_{i=1}^{I}c_{d}^{i}\mathcal{W}_{i}+\sum_{j=1}^{J}a_{d}^{j~\prime }%
\mathcal{D}_{j}\right] .  \label{trace}
\end{eqnarray}
Here $\mathcal{W}_{0}$ is the unique term of the form $W_{\mu \nu \rho
\sigma }\Box ^{n-2}W^{\mu \nu \rho \sigma }+\cdots $ such that $\sqrt{g}%
\mathcal{W}_{0}$ is conformally invariant, where $W^{\mu \nu \rho \sigma }$
is the Weyl tensor and the dots denote cubic terms in the curvature tensors.
Its coefficient $c_{d}$ is normalized so that for free fields ($n_{s}$ real
scalars, $n_{f}$ Dirac fermions and, in even dimensions, $n_{v}$ ($n-1$%
)-forms) it reads 
\begin{equation}
c_{d}=n_{s}+2^{[n]-1}(d-1)n_{f}+\frac{d!}{2\left[ \left( n-1\right) !\right]
^{2}}n_{v}.  \label{freec}
\end{equation}
The $\mathcal{D}_{j}$s, with $j>0$, are at least quadratic in the curvature
tensors.

The three central charges $c_{d}$, $a_{d}$, $a_{d}^{\prime }$ are special
among the others. They provide useful tools to investigate the properties of
conformal field theories in even dimensions and the space of
renormalization-group flows. In four dimensions, the trace anomaly in
external gravity at criticality contains only these three quantities: 
\[
\Theta _{*}=\frac{1}{120}{\frac{1}{(4\pi )^{2}}}\left[ c\,W^{2}-{\frac{a}{4}}%
\,\mathrm{G}+{\frac{2}{3}}\,a^{\prime }\,\Box R\right] . 
\]

\bigskip

The central charges satisfy positivity conditions ($c\geq 0$, $a\geq 0$),
encode the irreversibility of the RG$\;$flow ($\Delta a=a_{\mathrm{UV}}-a_{%
\mathrm{IR}}\geq 0$) and are sometimes computable exactly in the strongly
coupled IR limits of UV-free theories \cite{noi}. In this paper I\ discuss
several issues about the irreversibility of the RG\ flow in even and odd
dimensions and its relation with the central charges and the trace anomaly
in external gravity.

The formula (\ref{trace}) encodes a universal way to fix the relative
normalizations of $c$, $a$ and $a^{\prime }$ \cite{proc}, based on the
``pondered'' Euler density of ref. \cite{at6d} and the specialness of the $%
c=a$ theories of ref. \cite{cea}. The existence of a canonical relative
normalization for $c$, $a$ and $a^{\prime }$ makes it natural to inquire
whether the central charges are constrained by universal inequalities such
as $c\geq a$, $c\geq a^{\prime }$, $\Delta c\geq \Delta a^{\prime }$, etc.

Consider for example free field theories. The values of $c$ and $a$ in four
dimensions are 
\[
c=n_{s}+6n_{f}+12n_{v},\qquad a=\frac{1}{3}\left(
n_{s}+11n_{f}+62n_{v}\right) , 
\]
where $n_{s}$, $n_{f}$ and $n_{v}$ are the numbers of real scalars, Dirac
fermions and vectors, respectively. Note that scalars and fermions have $c>a$%
, while vectors have $c<a$. Analogous inequalities hold in higher even
dimensions. The equality $c=a$ holds in two dimensions. Is there a reason
why scalars and fermions always have $c\geq a$? Why vectors (or differential
forms, in higher dimensions) behave differently from scalars and fermions?
The answers to these questions are unexpectedly related to the
irreversibility of the RG flow.

The example just discussed is sufficient to exclude universal inequalities
between $c$ and $a$. For our purposes, it is more interesting to investigate
possible inequalities between $\Delta c$, $\Delta a$ and $\Delta a^{\prime }$%
. In an RG\ flow, the quantities $\Delta c$, $\Delta a$ and $\Delta
a^{\prime }$ have different properties: $\Delta a^{\prime }$ is non-negative
in unitary flows, $\Delta a$ is non-negative in unitary renormalizable
flows, while $\Delta c$ can be either positive or negative \cite{noi}.
Secondly, $\Delta c$ and $\Delta a$ are flow invariants of the second type 
\cite{fest}, which means that they depend only on the critical limits of a
flow and not on the particular flow connecting them. The reason is that the
central charges $c$ and $a$ are unambiguously defined at the critical points
by the embedding in external gravity. On the other hand, $a^{\prime }$ is
ill-defined at criticality and depends on the renormalization scheme.
Consequently, $\Delta a^{\prime }$ does depend on the particular flow
connecting the fixed points, although it is scheme-independent (the scheme
dependence cancels out in the difference $a_{\mathrm{UV}}^{\prime }-a_{%
\mathrm{IR}}^{\prime }$). The flow non-invariance of $\Delta a^{\prime }$ is
crucial for the ideas proposed below. Finally, $\Delta a^{\prime }$ has a
nice geometrical interpretation: it is the scheme-invariant area of the
graph of the (effective) beta function $\beta _{\mathrm{eff}}$ between the
fixed points \cite{athm}, where $\beta _{\mathrm{eff}}(|x|\mu )=|x|^{d}\sqrt{%
\langle \Theta (x)\,\Theta (0)\rangle }$, $\Theta $ being the trace of the
stress tensor and $d$ the space-time dimension. In classically conformal
theories $\Delta a^{\prime }$ is precisely the area of the beta function.

Some inequalities between $\Delta c$, $\Delta a$ and $\Delta a^{\prime }$
can be excluded immediately, using known results. In particular, the exact
formulas for $\Delta c$ and $\Delta a$ in supersymmetric theories derived in
refs. \cite{noi,noi2} allow us to compare $\Delta c$ and $\Delta a$ in a
variety of models and prove that no general inequality between $\Delta c$ to 
$\Delta a$ can hold. Combining the results of \cite{noi,noi2} with those of 
\cite{athm} it is possible to exclude also inequalities between $\Delta c$
and $\Delta a^{\prime }$. The calculations of \cite{noi,noi2}, however, do
not provide results for $\Delta a^{\prime }$. The present knowledge does not
allow us to exclude that a universal inequality might relate $\Delta a$ to $%
\Delta a^{\prime }$. This fact inspires some stimulating ideas.

\bigskip

It was recalled above that $\Delta a^{\prime }$ is non-negative in unitary
flows and does depend on the flow connecting the fixed points. This suggests
that $\Delta a^{\prime }$ measures the ``length'' of the flow. We have also
said that $\Delta a$ is non-negative in unitary renormalizable flows and
does not depend on the particular flow connecting the fixed points. This
suggests that $\Delta a$ measures the ``distance'' between the fixed points.
If this is correct, $\Delta a$ should be the minimum of $\Delta a^{\prime }$
in the space of flows connecting the same fixed points. It follows, in
particular, that unitary renormalizable flows satisfy the universal
inequalities $0\leq \Delta a\leq \Delta a^{\prime }$ in even dimensions.
This implies the irreversibility of the RG flow. The classically conformal
theories and the flows with $\Delta c=\Delta a$ are the ``geodesic flows'',
because they have $\Delta a=\Delta a^{\prime }$ \cite{athm,cea}.

The understanding I offer in this paper starts from these observations. It
includes and generalizes previous ideas about the irreversibility of the RG
flow in classically conformal theories and in the flows with $\Delta
a=\Delta c$, recently reviewed in ref. \cite{proc}. Among the other things,
I explain why scalars and fermions have necessarily $c\geq a$ and why
vectors can have $c<a$.

\bigskip

Several ideas apply also to odd-dimensional quantum field theory. The area $%
\Delta a^{\prime }$ of the graph of the effective beta function between the
fixed points can be taken as definition of length of the flow also in odd
dimensions. Its minimum over the flows $\digamma $ connecting the same UV\
and IR\ fixed points defines the distance between them, and the distance is
oriented. Moreover, it is still reasonable to expect that the flows with
minimal length are classically conformal. However, in odd dimensions the
minimum of $\Delta a^{\prime }$ cannot be expressed as the difference $%
\Delta a$ between the values of a central charge $a$ unambiguosly and
globally defined at criticality \cite{fest}. However, it is possible to give
a set of simple axioms that imply irreversibility also in odd dimensions.

The flows can be conventionally oriented from the UV to the IR. The distance
between the fixed points is therefore ``oriented'' and can satisfy certain
``oriented-triangle inequalities'' that are more restrictive than the usual
triangle inequalities. The notion of oriented distance and the
oriented-triangle inequalities are the basic axioms of irreversibility in
odd (and even) dimensions. First, they imply the irreversibility of the RG
flow without a ``height'' ($c$ or $a$) function. Irreversibility without an $%
a$ function is defined as the property that, in the realm of unitary
theories, there exist no pairs of non-trivial flows connecting interchanged
UV and IR fixed points. Second, the oriented-triangle inequalities imply the
existence of a local $a$ function, in the smooth regions of the space of
flows. Only in even dimensions there exists a global $a$ function.

In support of irreversibility without a global $a$ function in odd
dimensions, I check the oriented-triangle inequalities explicitly in a class
of three-dimensional flows introduced in ref. \cite{largeN}. In these
classically conformal models, the RG flow is exactly integrable at each
order of the large $N$ expansion \cite{largeN2}.

\bigskip

The ideas of this paper might have applications also outside quantum field
theory. Issues concerning the central charges $c$ and $a$ and the
irreversibility of the RG\ flow have recently attracted an amount of
interest in the realm of the so-called AdS/CFT correspondence \cite{malda},
a conjecture that relates the strongly coupled large N limits (where N is
typically a number of colors) of certain conformal field theories and RG
flows to supergravity and string duals. In particular, the conformal
theories considered in the AdS/CFT\ correspondence belong to the special
class of theories that have $c=a$ \cite{henningson}, at least to the leading
order in the large N expansion. These theories are mathematically elegant,
relatively simple, and have various properties in common with
two-dimensional conformal field theory \cite{cea}. In the context of the
AdS/CFT conjecture it is possible to derive the irreversibility of the RG
flow \cite{holoflow} for holographic flows interpolating between fixed
points belonging to the class $c=a$. This is the maybe best generalization
of Zamolodchikov's $c$ theorem \cite{zamolo}. I hope that the notions of
length of the RG flow, oriented distance between the fixed points and the
oriented-triangle inequalities might inspire new investigations in the
program of holographic renormalization (see \cite{holorew} for an account of
recent developments and references). In particular, it would be interesting
to have definitions of length of the flow and distance between the endpoints
of the flow in the context of string theory. The AdS/CFT\ conjecture is
certainly the best tool to start a research in this direction.

\bigskip

The paper is organized as follows. In section 2 I collect the main no-go
statements which exclude certain inequalities for trace anomalies. In
section 3 I inspect the sum rules for $a$ and $a^{\prime }$ to look for
inequalities between $\Delta a$ and $\Delta a^{\prime }$. I show that the
sum rules are compatible with the statements of this paper, but do not
provide an easy way to prove them. In section 4 I\ collect some observations
on irreversibility, which inspired the ideas of this paper. In section 5 I
formulate the definitions of length of the RG flow and distance between the
fixed points in even dimensions, relate the distance to $\Delta a$ and use
this relation to derive irreversibility in even dimensions. In section 6 I
analyse the main topological and metric properties of the spaces of fixed
points and flows. In section 7 I check some predictions in the case of free
fields. I explain why free scalars and fermions have $a\leq c$ in arbitrary
even dimensions, while free vectors can have $a>c$. I analyse also Gaussian
non-unitary fields, whose central charges $a$ and $c$ do not obey any
general inequality. In sections 8 and 9 I generalize the ideas to odd
dimensions. In section 8 I formulate the irreversibility of the RG flow
without an $a$ function and show that a good axiomatic set-up for
irreversibility is provided by the notion of oriented distance and the
oriented-triangle inequalities. I prove that the oriented-triangle
inequalities imply irreversibility without a global $a$ function. In section
9 I test the oriented-triangle inequalities in a class of three-dimensional
classically conformal flows. Section 10 contains the conclusions.

\section{No-go statements}

\setcounter{equation}{0}

In this section I use known results to exclude some inequalities between the
trace anomalies. I recall that there exist classes of flows with $\Delta
c=\Delta a$ \cite{cea}, with $\Delta c=\Delta a^{\prime }$ \cite{inv} and
with $\Delta a=\Delta a^{\prime }$ \cite{athm}. Here I prove that no general
inequalities of the form $\Delta c\geq \leq f_{1}\Delta a$, $\Delta c\geq
\leq f_{2}\Delta a^{\prime }$ hold, where $f_{1,2}$ are numerical factors.
There survives only the possibility of a general inequality relating $\Delta
a$ and $\Delta a^{\prime }$.

\medskip

\textbf{No $\Delta c-\Delta a$ inequality.} In a class of supersymmetric
theories in the conformal window the values of $\Delta a$ and $\Delta c$
have be computed exactly \cite{noi,noi2}. For our considerations, it is
sufficient to consider the examples treated explicitly in ref. \cite{noi},
in particular an ``electric'' theory and its dual ``magnetic'' theory. Their
IR conformal fixed points are related by Seiberg's ``electric-magnetic''
duality \cite{seiberg}.

The electric theory is N=1 supersymmetric QCD with $N_{c}$ colors and $N_{f}$
flavors. In the conformal window $3N_{c}/2\leq N_{f}\leq 3N_{c}$ the results
are 
\begin{eqnarray*}
\Delta a &=&{\frac{5}{2}}N_{c}N_{f}\left( 1-3{\frac{N_{c}}{N_{f}}}\right)
^{2}\left( 2+3{\frac{N_{c}}{N_{f}}}\right) ,\qquad \Delta c={\frac{5}{2}}%
N_{c}N_{f}\left( 1-3{\frac{N_{c}}{N_{f}}}\right) \left( 4-3{\frac{N_{c}}{%
N_{f}}}-9{\frac{N_{c}^{2}}{N_{f}^{2}}}\right) , \\
&&\qquad \qquad \qquad \qquad \qquad \frac{\Delta c}{\Delta a}=\frac{%
4-3r-9r^{2}}{(1-3r)(2+3r)},
\end{eqnarray*}
where $r=N_{c}/N_{f}$. We see that in the conformal window $\Delta c/\Delta
a $ ranges from $-\infty $ to $1/2$.

The magnetic theory is N=1 supersymmetric QCD with $N_{c}$ colors, $N_{f}$
quark flavors and a $N_{f}\times N_{f}$ meson superfield. The IR fixed point
of the magnetic theory is related to the IR fixed point of the electric
theory by means of the duality map $N_{c}^{\mathrm{mag}}=N_{f}^{\mathrm{el}%
}-N_{c}^{\mathrm{el}}$ and $N_{f}^{\mathrm{mag}}=N_{f}^{\mathrm{el}}$ \cite
{seiberg}. In particular, duality relates the IR values of $c$ and $a$ in
the electric and magnetic models. Instead, the UV values of $c$ and $a$ (and
therefore the differences $\Delta c$, $\Delta a$) are not related to each
other. The magnetic theory provides an independent source of information
about possible inequalities for $\Delta c$ and $\Delta a$.

In the conformal window the results are 
\begin{eqnarray}
\Delta a &=&{\frac{5}{2}}\left( 1-3{\frac{N_{c}}{N_{f}}}\right) ^{2}\left(
3N_{c}^{2}-10N_{c}N_{f}+10N_{f}^{2}\right) ,  \nonumber \\
\Delta c &=&{\frac{5}{2}}\left( 1-3{\frac{N_{c}}{N_{f}}}\right) \left(
8N_{f}^{2}-38N_{c}N_{f}+33N_{c}^{2}-9{\frac{N_{c}^{3}}{N_{f}}}\right) , 
\nonumber \\
\frac{\Delta c}{\Delta a} &=&\frac{10-10r+3r^{2}}{%
(1-3r)(8-38r+33r^{2}-9r^{3})}.  \label{see}
\end{eqnarray}
Here $\Delta c/\Delta a$ ranges from $7/8$ to $+\infty $, which has no
overlap with the range of $\Delta c/\Delta a$ in the electric theory. From
these two examples, it is immediate to conclude that no general inequality
can relate $\Delta a$ and $\Delta c$.

\medskip

\textbf{No $\Delta c-\Delta a^{\prime }$ inequality.} In classically
conformal theories, we have $\Delta a=\Delta a^{\prime }$ to the fourth loop
order in perturbation theory and an argument saying that the relation $%
\Delta a=\Delta a^{\prime }$ is actually exact \cite{athm}. Here we just
need to know that $\Delta a^{\prime }\sim (1-3r)^{2}$ when $r$ tends to $1/3$%
. I recall that in the limit $r\gtrsim 1/3$, the perturbative expansion can
be used, since the IR fixed point is weakly coupled. Moreover, $\Delta
a^{\prime }$ is always non-negative, because of the sum rule (\ref{magna})
(see below). Now, in the electric theory, when $r$ gets close to $1/3$ the
ratio $\Delta c/\Delta a^{\prime }$ takes arbitrarily large negative values.
Instead, in the magnetic theory when $r$ gets close to $1/3$ the ratio $%
\Delta c/\Delta a^{\prime }$ takes arbitrarily large positive values. This
proves that there exists no universal inequality relating \textbf{$\Delta c$}
to \textbf{$\Delta a^{\prime }$}.

\medskip

After this analysis, we remain only with the possibility of an inequality
for $\Delta a$ and $\Delta a^{\prime }$. The methods of refs. \cite{noi,noi2}
do not allow us to calculate $\Delta a^{\prime }$ in the supersymmetric
conformal windows and compare it to $\Delta a$.

\section{Sum rules in even dimensions}

\setcounter{equation}{0}

In even dimension greater than two the embedding in external gravity allows
us to derive sum rules for $\Delta a$ and $\Delta a^{\prime }$ \cite{234}.
It is instructive to inspect these sum rules to see whether we can prove a
general inequality for $\Delta a$ and $\Delta a^{\prime }$ .

In arbitrary even $d=2n$ dimensions the sum rule for $\Delta a^{\prime }$ is 
\cite{at6d} 
\begin{equation}
\Delta a^{\prime }=\frac{\pi ^{n}\,(d+1)}{{n}!}\int \mathrm{d}%
^{d}x\,|x|^{d}\,\langle \Theta (x)\,\Theta (0)\rangle .  \label{magna}
\end{equation}
By definition, the contact term is excluded from the integral. The integral (%
\ref{magna}) is over all points of spacetime and depends, in general, on the
parameters of the theory (ratios between masses, values of the coupling
constants at a reference energy, etc.).

When the operator $\Theta $ is finite, which happens for example in
four-dimensional theories containing no scalars (when scalar fields are
present $\Theta $ mixes with an improvement term for the stress tensor and
the treatment is more complicated \cite{inv}), it is possible to parametrize
the correlation function $\langle \Theta (x)\,\Theta (0)\rangle ,$ using the
Callan-Symanzik equation, in terms of a function $G$ of the running coupling
constants $g$ at the scale $1/|x|$: 
\[
\langle \Theta (x)\,\Theta (0)\rangle =\frac{G(g(1/|x|))}{|x|^{2d}}\equiv 
\frac{\widetilde{G}(g(\mu ),\ln |x|\mu )}{|x|^{2d}}. 
\]
Then the integral (\ref{magna}) becomes an integral over the RG flow: 
\[
\Delta a^{\prime }=\frac{d(d+1)}{(n!)^{2}}\pi ^{d}\int_{-\infty }^{+\infty }%
\mathrm{d}t\,~\widetilde{G}(g(\mu ),t). 
\]
The behavior of $\widetilde{G}$ for large $|t|$ can be studied using the RG\
equations. The integral is ensured to be convergent, if the flow
interpolates between well-defined UV and IR limits.

\bigskip

In the realm of unitary theories we have $\Delta a^{\prime }\geq 0$, because
reflection positivity ensures $\langle \Theta (x)\,\Theta (0)\rangle \geq 0$
for $|x|\neq 0$. The equality $\Delta a^{\prime }=0$ takes place only if the
RG flow is trivial ($\Theta =0$), i.e. the UV\ and IR fixed points coincide
or belong to a family of continuously connected conformal field theories.

The formula for $\Delta a$ is the sum of the integral (\ref{magna}) plus
integrals of correlation functions containing more $\Theta $-insertions. For
example, in four dimensions two equivalent sum rules for $\Delta a$ read 
\cite{234} 
\begin{eqnarray}
\Delta a &=&-{5\frac{\pi ^{2}}{2}}\int \mathrm{d}^{4}x\,|x|^{4}\,\Gamma
_{x0}^{\prime }-{5\frac{\pi ^{2}}{2}}\int \mathrm{d}^{4}x\,\mathrm{d}%
^{4}y\,x^{2}\,y^{2}\,\Gamma _{xy0}^{\prime }  \label{ti} \\
&=&-{\frac{5\pi ^{2}}{2}}\int \mathrm{d}^{4}x\,|x|^{4}\,\Gamma _{x0}^{\prime
}-{\frac{5\pi ^{2}}{2}}\int \mathrm{d}^{4}x\,\mathrm{d}^{4}y\,\mathrm{d}%
^{4}z\,\left( x\cdot y\right) \left( x\cdot z\right) \Gamma _{xyz0}^{\prime
},  \nonumber
\end{eqnarray}
while in six dimensions the ``minimal'' sum rule reads \cite{fest} 
\begin{eqnarray}
\Delta a &=&\frac{7\pi ^{3}}{36}\left( -6\int x^{6}\;\Gamma _{x,0}^{\prime
}\;\mathrm{d}^{6}x+\int [8(x\cdot y)^{3}-9x^{4}y^{2}]\;\Gamma
_{x,y,0}^{\prime }\;\mathrm{d}^{6}x\mathrm{d}^{6}y\right.  \nonumber \\
&&\qquad \qquad \qquad \qquad \qquad \left. +6\int x^{2}(y\cdot
z)^{2}\;\Gamma _{x,y,z,0}^{\prime }\;\mathrm{d}^{6}x\mathrm{d}^{6}y\mathrm{d}%
^{6}z\right) .  \label{minimale}
\end{eqnarray}
Here $\Gamma $ is the quantum action in external gravity, $\Gamma ^{\prime
}=\Gamma -\Gamma _{\mathrm{UV}}$ and $\Gamma _{x_{1}\cdots x_{k}}$ is the $k$%
th functional derivative of $\Gamma $ with respect to the conformal factor $%
\phi $ at the points $x_{1}\cdots x_{k}$. For example, $\Gamma _{x0}^{\prime
}=-\langle \Theta (x)$\ $\Theta (0)\rangle $, etc. In arbitrary even
dimension $2n$ we have integrals of correlation functions containing up to $%
n+1$ insertions of $\Theta $.

It does not seem straightforward to derive general inequalities between $%
\Delta a$ and $\Delta a^{\prime }$ using the sum rules just written, because
there is no simple way \cite{234} to apply Osterwalder-Schrader (OS)
positivity \cite{os}. However, we can make some observations. Using the
vanishing relations of \cite{234} an equivalent form for the
four-dimensional $\Delta a$ sum rules (\ref{ti})can be written, namely 
\begin{equation}
\Delta a=\Delta a^{\prime }+{\frac{\pi ^{2}}{96}\lim_{V\rightarrow \infty }}%
\frac{1}{V}\int_{V}\mathrm{d}^{4}x\,\mathrm{d}^{4}y\,\mathrm{d}^{4}z\,%
\mathrm{d}^{4}w\,\left( x-y\right) ^{2}\left( z-w\right) ^{2}\Gamma
_{xyzw}^{\prime }.  \label{basic}
\end{equation}
Here the integrand is a positive function times $\Gamma _{xyzw}^{\prime }$.
Basically, $\Gamma _{xyzw}^{\prime }$ is minus the $\Theta $ four-point
function. Naively, it is tempting to think that the four-point function is
``positive'', in some sense. This suggests that the integral in (\ref{basic}%
) is negative or zero and the inequality 
\[
\Delta a^{\prime }\geq \Delta a 
\]
holds.

The argument is however naive, for the reasons that I\ now explain. OS
positivity states that the integral 
\begin{equation}
\int \mathrm{d}^{4}x\,\mathrm{d}^{4}y\,\mathrm{d}^{4}z\,\mathrm{d}^{4}w\
g(x,y)\,g^{*}(\theta z,\theta w)\,\left\langle \mathrm{O}(x)\ \mathrm{O}(y)\,%
\mathrm{O}(z)\ \mathrm{O}(w)\right\rangle  \label{legi}
\end{equation}
is non-negative, for every Hermitean operator O and function $g(x,y)$,
vanishing together with its derivatives unless $x^{0}>y^{0}>0$. Here $\theta
(x^{0},x^{1},x^{2},x^{3})=(-x^{0},x^{1},x^{2},x^{3})$. The positivity
condition holds for every choice of the ``time'' axis $x^{0}$.

The application of OS positivity to our integral (\ref{basic}) is
problematic, however. The function $g(x,y)$ is $\left( x-y\right) ^{2}/\sqrt{%
V}$ inside the finite volume $V$ and zero elsewhere, so it does not vanish
together with its derivatives if $x^{0}>y^{0}>0$ is not true.

Due to the symmetry of the correlation function under the exchange of $x$
and $y$, the requirement that $g(x,y)$ should vanish together with its
derivatives unless $x^{0}>y^{0}$ can be replaced with the requirement that $%
g(x,y)$ should vanish together with its derivatives at $x^{0}=y^{0}$. As
long as $x\neq y$, there is no reason to expect surprises at $x^{0}=y^{0}$,
but we do have to pay attention to the coincident points (see below).

I now show that if the points $x$, $y$, $z$, $w$ do not lie on a plane, we
can also relax the restriction that the product $g(x,y)g^{*}(\theta z,\theta
w)$ should vanish together with its derivatives unless $%
x^{0},y^{0}>0>z^{0},w^{0}$. It is easy to prove, using the invariance under
translations and rotations, a simple theorem, stating that for every set of
points $x$, $y$, $z$ and $w$ that do not lie on a plane, there exist an
origin and a time axix such that $x^{0},y^{0}>0>z^{0},w^{0}$ with repect to
that origin and that axis. When $x$, $y$, $z$ and $w$ do lie on a plane (in
particular, when two points coincide), there might exist no origin and time
axis such that $x^{0},y^{0}>0>z^{0},w^{0}$. Now, the function $g(x,y)$ is
invariant under translations and rotations, if we neglect that the integrals
of (\ref{basic}) are performed in a finite volume $V$. Since we have to take
the limit $V\rightarrow \infty $ in the end, it is probably legitimate to
ignore this nuisance. Then, whenever the points $x$, $y$, $z$ and $w$ do not
lie on a plane we can apply our simple theorem to (\ref{legi}) and choose an
origin and a time axis such that OS\ positivity holds. So, the contributions
to the integral (\ref{legi}) coming from distinct points appear to be under
control.

Ultimately, the true difficulty to apply OS positivity to the integral (\ref
{basic}) comes from the coincident points. The set of coincident points is
of vanishing measure only if the correlation functions have no contact
terms. If there are no contact terms, we can surround the points $x$, $y$, $%
z $ and $w$ with infinitesimal spheres, perform the integral outside the
spheres and let the radii of the spheres tend to zero in the very end. If
there are contact terms, however, this procedure does not return the correct
result and we cannot conclude that the integral (\ref{legi}) satisfies
positivity.

There is no reason to expect that contact terms are absent. Actually, simple
perturbative calculations show that contact terms are expected to be there.
In momentum space, for example, concact terms are the product of a local
function of some momenta times an arbitrary function of the other momenta.
Finally, the contact terms of a four-point function are associated with
three- or two-point functions. We have no control on the positivity of the
flow integrals of the three-point functions.

Moreover, in (\ref{basic}) we do not just have the $\Theta $ four-point
function, but $-\Gamma _{xyzw}^{\prime }$, which is a combination of four-,
three- and two-point functions \cite{234}. The difference between $-\Gamma
_{xyzw}^{\prime }$ and the $\Theta $ four-point function is made of other
contact terms.

Having shown that it is illegitimate to ignore the contact terms, we cannot
rigorously prove the inequality $\Delta a^{\prime }\geq \Delta a$. On the
other hand, this difficulty is more than welcome. If the contact terms were
not there, the above arguments would imply the strict inequality $\Delta
a^{\prime }>\Delta a$ any time the flow is nontrivial ($\Theta \neq 0$).
This would contraddict the claim of ref. \cite{athm} that in classically
conformal theories the equality $\Delta a^{\prime }=\Delta a$ holds exactly.
Moreover, I would not be allowed to argue, as I do below, that the minimum
of $\Delta a^{\prime }$ over the flows $\digamma $ connecting the same fixed
points is precisely $\Delta a$.

In conclusion, the inequality $\Delta a^{\prime }\geq \Delta a$ is possible,
and there is room for nontrivial flows satisfying the equality $\Delta
a^{\prime }=\Delta a$. Actually, if we enlarge the class of flows to include
nonrenormalizable (e.g. asymptotically safe) theories, then there is also
room for flows with $\Delta a<0$ (see below).

\section{Irreversibility of the RG flow in even dimensions}

\setcounter{equation}{0}

In even dimension greater than two and in odd dimensions the embedding in
external gravity is unable to explain many known properties of trace
anomalies \cite{fest}. Some properties, verified empirically in a number of
cases, remain unexplained. To achieve a better understanding, it is helpful
to investigate the space of conformal field theories and RG flows connecting
them, and study the topological and metric properties of this space.

In even dimensions, the irreversibility of the RG flow \cite{zamolo,proc} is
expressed by the existence of a positive quantity $a$ whose values are
always larger in the ultraviolet than in the infrared. The quantity $a$ is
interpreted as a counter of the massless degrees of freedom of the theory.
In two dimensions, this quantity is Zamolodchikov's $c$ function. In higher
even dimensions it is the central charge $a$, namely the coefficient of the
Euler density in the trace anomaly of the theory embedded in external
gravity.

In refs. \cite{athm,at6d,cea} an approach to the irreversibility of the RG
flow in even dimensions has been developed. A synthetic review can be found
in \cite{proc}. The approach of these references does not apply to the most
general flow, but only the subclass of classically conformal flows and the
subclass of flows that have $\Delta a=\Delta c$. Relevant and irrelevant
deformations have to be included in a more complete understanding that
possibly applies also to odd dimensions.

In this section I collect some considerations about the irreversibility of
the RG flow in even dimensions, which hopefully make the inequality $\Delta
a^{\prime }\geq \Delta a$ more pausible. I stress the different roles played
by classically conformal and classically non-conformal theories.

In quantum field theory, there are basically two sources of violations of
scale invariance: the dimensionful parameters of the classical lagrangian
and the dynamical scale $\mu $ introduced by renormalization. Power counting
groups the dimensionful parameters into relevant and irrelevant. The
relevant parameters are expected to enhance irreversibility, because of the
Appelquist-Carazzone decoupling theorem \cite{appel}. For example, the
theories of massive free scalars and fermions certainly have $a_{\mathrm{UV}%
}>a_{\mathrm{IR}}$. Symmetrically, the irrelevant parameters are expected to
depress irreversibility, because a non-renormalizable coupling, which does
not affect the IR limit, can kill degrees of freedom in the UV limit.
Observe that for these considerations, which mostly concern the signs of $%
\Delta a$, $\Delta a^{\prime }$ and $\Delta a-\Delta a^{\prime }$, it is
necessary to assume only that the theory is unitary and interpolates between
well-defined UV and IR fixed points. In particular, it is not necessary to
assume that the theory is renormalizable in a conventional sense, nor that
it is predictive, i.e. quantizable with finitely many parameters. In the
realm of non-renormalizable theories, consistent flows can be defined, for
example, in the scenario of Weinberg's asymptotic safety \cite{wein}.

In view of the observations just made, the irrelevant parameters, such as
the Newton constant, are expected to violate the irreversibility of the RG
flow ($\Delta a<0$), unless they are dynamically generated by the
renormalization scale $\mu $ (from a renormalizable theory). Note that the
existence of flows with $\Delta a<0$ does not contraddict irreversibility,
as long as $\Delta a<0$ is the product of an \textit{explicit} violation of
scale invariance, that is to say a violation due to a classical dimensionful
parameter. It is not surprising that explicit violations produce $\Delta a<0$%
, because they can be arbitrarily strong and cover all opposite effects. The
theories that contain no explicit violation of scale invariance are
precisely the classically conformal theories. There, the dynamical breaking
of scale invariance does not mix with the effects of explicit violations.
This is the reason why the classically conformal theories occupy a special
role in the investigation of irreversibility.

The properties of relevant, marginal and irrelevant parameters at the
classical and quantum levels are summarized in the table 
\[
\begin{tabular}{|c|c|c|}
\hline
& classic & quantum \\ \hline
relevant & $\Delta a^{\prime }>0,\quad \Delta a>0,$ & $\Delta a^{\prime
}>0,\quad \Delta a>0,$ \\ \hline
marginal & $\Delta a^{\prime }=0,\quad \Delta a=0,$ & $\Delta a^{\prime
}\geq 0,\quad \Delta a\geq 0,$ \\ \hline
irrelevant & $\Delta a^{\prime }>0,\quad \Delta a<0,$ & $\Delta a^{\prime
}>0,\quad \Delta a><0.$ \\ \hline
\end{tabular}
\]
At the classical level, the marginal plane $\Delta a=0$ separates the space
of relevant flows ($\Delta a>0$) from the space of irrevelant flows ($\Delta
a<0$). At the quantum level, the plane $\Delta a=0$ moves inside the space
of irrelevant flows. Now, $\Delta a^{\prime }$ is strictly positive in all
non-trivial unitary theories, but asymptotically safe flows with $\Delta
a\leq 0$ (and $\Delta a^{\prime }>0$) are in principle allowed to exist.
These considerations rule out the inequality $\Delta a^{\prime }\leq \Delta
a $.

In conclusion, only one universal inequality is not ruled out, namely 
\begin{equation}
\Delta a^{\prime }\geq \Delta a.  \label{ine}
\end{equation}
This possibility opens the door to some stimulating ideas.

\section{Length of the RG\ flow, distance between the fixed points and $%
\Delta a$}

In this section I elaborate the definitions of length of the RG\ flow and
oriented distance between the fixed points. The RG\ flow is conventionally
oriented from the ultraviolet to the infrared. I conjecture that the
distance coincides with $\Delta a$. The inequalities $\Delta a^{\prime }\geq
\Delta a\geq 0$ and the irreversibility of the RG flow (in even dimensions)
are implied straightforwardly by this conjecture. The generalization of
these ideas to odd dimensions is presented in section 8.

\bigskip

\textbf{Length of the RG flow and distance between the fixed points.} The
quantity $\Delta a^{\prime }$ is always positive and does depend on the flow
connecting the fixed points. It is therefore a natural candidate to define
the length $L$ of the flow $\digamma $: 
\begin{equation}
L(\digamma )=\frac{\pi ^{n}\,(d+1)}{{n}!}\int_{\digamma }\mathrm{d}%
^{d}x\,|x|^{d}\,\langle \Theta (x)\,\Theta (0)\rangle =\Delta a^{\prime
}(\digamma ).  \label{length}
\end{equation}
A (unitary) flow of zero length is trivial, since $L(\digamma )=0$ and
reflection positivity imply $\Theta \equiv 0$.

The minimum of $L(\digamma )$ in the space $\mathcal{F}_{\mathrm{%
C_{UV},C_{IR}}}$ of (unitary, renormalizable) flows $\digamma $ connecting
the same fixed points C$_{\mathrm{UV}}$ and C$_{\mathrm{IR}}$ is the
distance $d$ between them: 
\begin{equation}
d(\mathrm{C}_{\mathrm{UV}},\mathrm{C}_{\mathrm{IR}})=\min_{\digamma \in 
\mathcal{F}_{\mathrm{C_{UV},C_{IR}}}}L(\digamma ).  \label{distance}
\end{equation}
The minimum has to be taken in the space of continuously deformable flows
and sequences of flows with concordant orientations. If the space $\mathcal{F%
}_{\mathrm{C_{UV},C_{IR}}}$ has disconnected components, there might be a
minimum in each subspace $\mathcal{F}_{1}$, $\mathcal{F}_{2}$, $\ldots $ of
continuously connected flows. In some situations, $L(\digamma )$ might not
admit a minimum, but only an inferior limit. Then the distance is defined as
the inferior limit of $L(\digamma )$.

The distance between the conformal field theories from C$_{\mathrm{1}}$ and $%
\mathrm{C}_{\mathrm{2}}$ is defined only if there exists a flow, or a
sequence of flows with concordant orientations, interpolating between C$_{%
\mathrm{1}}$ and $\mathrm{C}_{\mathrm{2}}$.

The distance between two continuously connected conformal field theories
(e.g. two N=4 $d=4$ supersymmetric Yang-Mills theories with different values
of the gauge coupling $g$) is zero, because an exactly marginal deformation
of a conformal field theory is a trivial RG\ flow.

The minimization of $\Delta a^{\prime }$ in the space of flows was first
realized to have remarkable properties in ref. \cite{inv}.

With an abuse of language, I use the words ``distance between the fixed
points'', even if, strictly speaking, $d(\mathrm{C}_{\mathrm{UV}},\mathrm{C}%
_{\mathrm{IR}})$ is the distance between suitable projections of the fixed
points. The space of conformal theories $\mathcal{C}$ is projected to a
space $\pi \mathcal{C}$ and the projecton $\pi $ is such that conformal
field theories with zero distance are projeced onto the same point of $\pi 
\mathcal{C}$. Similarly, the space $\mathcal{F}$ is projected onto a space $%
\pi \mathcal{F}$.

\bigskip

\textbf{Oriented distance between the fixed points and $\Delta a$}. I now
restrict the attention to renormalizable theories, on which we have a better
control. Inspired by the considerations made so far, in particular $i$) the
possibility of a universal inequality (\ref{ine}), $ii$) the existence of
flows with $\Delta a^{\prime }=\Delta a$ (the classically conformal
theories) and $iii$) the independence of $\Delta a$ from the flow connecting
the fixed points, I conjecture that the distance between the fixed points is
precisely $\Delta a$, i.e. 
\begin{equation}
d(\mathrm{C}_{\mathrm{UV}},\mathrm{C}_{\mathrm{IR}})=\Delta a=a_{\mathrm{UV}%
}-a_{\mathrm{IR}}=\min_{\digamma \in \mathcal{F}_{\mathrm{C_{UV},C_{IR}}%
}}\Delta a^{\prime }(\digamma ).  \label{auto}
\end{equation}
The relation (\ref{auto}) encodes also the fact that the distance is
oriented. This means, in particular, that the distance is not symmetric:
strictly speaking $d(\mathrm{C}_{\mathrm{UV}},\mathrm{C}_{\mathrm{IR}})$ is
the distance \textit{from} $\mathrm{C}_{\mathrm{UV}}$ to $\mathrm{C}_{%
\mathrm{IR}}$, not the distance \textit{between} $\mathrm{C}_{\mathrm{UV}}$
and $\mathrm{C}_{\mathrm{IR}}$. The axioms of the oriented distance are
elaborated in section 8.

Furthermore, (\ref{auto}) allows us to conclude that $\pi \mathcal{C}$ and $%
\pi \mathcal{F}$ are one-dimensional, a subset of the $a$-axis.

Formula (\ref{auto}) implies the inequalities 
\begin{equation}
0\leq \Delta a\leq \Delta a^{\prime }  \label{ineq}
\end{equation}
and therefore the irreversibility of the RG flow as it is commonly stated in
even dimensions ($\Delta a\geq 0$).

The observations made in the previous section suggest also that if we extend
our considerations to non-renormalizable flows, e.g. asymptotically safe
theories, the minimum (\ref{auto}) has probably no relation with $\Delta a$.

The ideas of the present paper generalize the understading of ref.s \cite
{proc,athm,at6d,cea}, which mostly concerned classically conformal flows and
the flows with $\Delta c=\Delta a$. Several pieces of evidence suggested
that these flows have $\Delta a=\Delta a^{\prime }$, namely that they
saturate the minimum (\ref{auto}) (``geodesic'' flows). The extended picture
covers also non-geodesic flows, for which the strict inequality $\Delta
a^{\prime }>\Delta a$ can hold.

Observe that in two dimensions the three central charges $c$, $a$ and $%
a^{\prime }$ are indistinguishable at the fixed points, since the trace
anomaly in external gravity contains only one term, namely $cR/(24\pi )$.
With the relative normalization adopted in the introduction we are allowed
to write the identifications $c=a=a^{\prime }$ in two dimensions. This means
that all of the flows have $\Delta c=\Delta a=\Delta a^{\prime }$ and
therefore equal and minimal length.

In section 7 I derive some predictions from the statements of this section
and test them.

\section{Geometry of the spaces of fixed points and flows}

In this section I make some observations about the topological and metric
properties of the spaces $\mathcal{C}$ and $\mathcal{F}$ of fixed points and
flows in even-dimensional quantum field theory. This kind of analysis is
extended to odd-dimensional theories in section 8.

\begin{figure}[tbp]
\centerline{\epsfig{figure=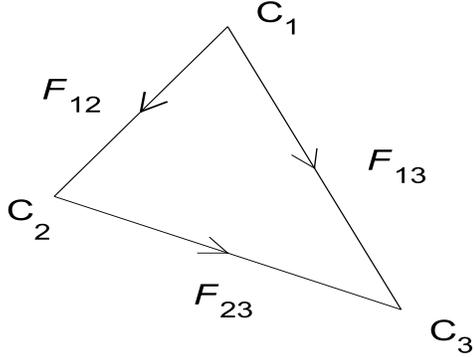,height=5cm,width=9cm}}
\caption{An oriented triangle in the space of flows}
\end{figure}
\bigskip

\textbf{All triangles are degenerate.} First I show that all triangles are
degenerate in the space $\pi \mathcal{C}$. Consider a ``triangle'' T in the
space of flows (Fig. 1), made of the fixed points C$_{1}$, C$_{2}$ and C$_{3}
$, and the RG flows $\digamma _{12}$, $\digamma _{23}$, and $\digamma _{13}$%
, connecting C$_{1}$ to C$_{2}$, C$_{2}$ to C$_{3}$ and C$_{1}$ to C$_{3}$,
respectively. Due to the existence of the central charge $a$ and the
relation (\ref{auto}), the distance $d_{ij}$ between C$_{i}$ and C$_{j}$ is
oriented from C$_{i}$ to C$_{j}$ and equal to $a_{i}-a_{j}$ with $i<j$. We
have the ``triangle equality'' 
\begin{equation}
d_{13}=a_{1}-a_{3}=(a_{1}-a_{2})+(a_{2}-a_{3})=d_{12}+d_{23}.  \label{treq}
\end{equation}
The geometric meaning of the triangle equality is that the space $\pi 
\mathcal{C}$ in which the distances are measured is one-dimensional.

The triangle equality is a property of even-dimensional quantum field
theory. Odd-dimensional quantum field theory does not admit a global $a$
function and $\pi \mathcal{C}$ can have dimension greater than one.

\bigskip

In higher even dimensions, the central charges $c$ and $a$ do not identify a
conformal field theory uniquely. There exist families of continuously
connected inequivalent conformal field theories having the same $c$ and the
same $a$ \cite{noi}. If we embed the space $\mathcal{C}$ of conformal field
theories in $\mathbf{R}^{k}$ for $k$ sufficiently large, we probably find
sets of discrete points, lines (one-parameter families of continuously
connected conformal field theories), two-dimensional surfaces and three- or
higher-dimensional regions. Probably, most theories are isolated points or
belong to one-parameter families, and the higher-dimensional regions are
exceptional. The space $\mathcal{F}$ is even more complicated, since it is
the space of flows connecting the points of $\mathcal{C}$. We do not know if
there exists a flow, or a family of flows, connecting every pair of points
of $\mathcal{C}$; several pairs of points might not admit a flow connecting
them. Some families of flows are continuously connected, others are not.
Probably, $\mathcal{F}$ looks like a neural network. Observe that the notion
of oriented distance works quite well for a neural network, or in general an
environment where the paths connecting the nodes are one-way and some pairs
of nodes are connected by no path nor sequence of paths with concordant
orientations.

The topology of $\pi \mathcal{C}$ is considerably simpler that the topology
of $\mathcal{C}$: $\pi \mathcal{C}$ is\ just a set of points and maybe
intervals on the $a$ axis. We can expect that also the topology of $\pi 
\mathcal{F}$ is simpler than the topology of $\mathcal{F}$.

\bigskip

Finally, regions of $\mathcal{C}$ with different values of $a$ or $c$ are
disconnected from one another. This follows from a property of the central
charges $c$ and $a$ known as ``marginality of the central charge'' \cite{noi}%
, stating that continuously connected conformal field theories have the same 
$c$ and $a$.

\section{Inequalities for trace anomalies in free-field theories}

\setcounter{equation}{0}

I have anticipated, in the introduction, that free scalars and fermions have 
\begin{equation}
a_{\mathrm{free}}\leq c_{\mathrm{free}}  \label{freeineq}
\end{equation}
in arbitrary even dimensions $d=2n$, while vectors, or, more generally, the $%
(n-1)$-differential forms, have $a\geq c$. The values of $c$ are \cite{iera} 
\[
c_{\mathrm{scal}}=1,\qquad c_{\mathrm{ferm}}=2^{n-1}(d-1),\qquad c_{\mathrm{%
forms}}={\frac{1}{2}}{\frac{d!}{[(n-1)!]^{2}}}, 
\]
while the values of $a$ have been calculated in \cite{cappelli}. Using the
procedure of section 2.3 of \cite{fest} we can write 
\[
a={\frac{1}{2}}(-1)^{n-1}(-1)^{2S}{\frac{(2n+1)!}{n!(n-1)!}}\Upsilon (0) 
\]
where $S$ is the spin and 
\[
\Upsilon (0)=\lim_{s\rightarrow 0}\sum_{k=0}^{\infty }{\frac{\delta _{k}}{%
\varpi _{k}^{s}}}. 
\]
Here $\varpi _{k}$ are the eigenvalues of an appropriate second-order
differential operator and $\delta _{k}$ are their multiplicities on the
sphere ($R_{\mu \nu }=\Lambda g_{\mu \nu }$). The differential operator is $%
-\Box +\Lambda d(d-2)/(4(d-1))$ for scalar fields and $-\Box +d\Lambda /4$
for fermions. In the case of the differential forms, the sum $\Upsilon (0)$
is defined as the $\zeta _{AT}(0)$ of \cite{cappelli}.

Using the values of $\Upsilon (0)$ given in the tables of \cite{cappelli},
we have, for scalars and fermions, 
\[
\begin{tabular}{|c|c|c|c|c|c|c|c|}
\hline
$d$ & 2 & 4 & 6 & 8 & 10 & 12 & 14 \\ 
${a_{\mathrm{scal}}/c_{\mathrm{scal}}}$ & $1$ & ${\frac{1}{3}}$ & ${\frac{5}{%
18}}$ & ${\frac{23}{90}}$ & ${\frac{263}{1080}}$ & ${\frac{133787}{567000}}$
& ${\frac{157009}{680400}}$ \\ 
${a_{\mathrm{ferm}}/c_{\mathrm{ferm}}}$ & $1$ & ${\frac{11}{18}}$ & ${\frac{%
191}{360}}$ & ${\frac{2497}{5040}}$ & ${\frac{14797}{31104}}$ & ${\frac{%
92427157}{199584000}}$ & ${\frac{36740617}{80870400}}$%
\end{tabular}
\]
etc. We see that (\ref{freeineq}) is satisfied. The ratios $a/c$ decrease
when the dimension increases.

Now consider the differential forms. The ratio $a/c$ is equal to 
\[
{\frac{a_{\mathrm{forms}}}{c_{\mathrm{forms}}}}=(-1)^{n-1}{\frac{d+1}{n}}%
\Upsilon (0). 
\]
Taking $\Upsilon (0)$ from the tables of \cite{cappelli}, we have 
\[
\begin{tabular}{|c|c|c|c|c|c|c|}
\hline
$d$ & 4 & 6 & 8 & 10 & 12 & 14 \\ 
${a_{\mathrm{forms}}/c_{\mathrm{forms}}}$ & ${\frac{31}{18}}$ & ${\frac{221}{%
90}}$ & ${\frac{8051}{2520}}$ & ${\frac{1339661}{340200}}$ & ${\frac{%
525793111}{112266000}}$ & ${\frac{3698905481}{681080400}}$%
\end{tabular}
\]
etc. The free $(n-1)$-differential forms do not satisfy (\ref{freeineq}).
Actually, they always have $a>c$.

\bigskip

I now prove that the inequality (\ref{freeineq}) for scalars and fermions
follows from the inequality (\ref{ineq}) of section 5. Consider the RG\
flows of free massive scalars and fermions. In the ultraviolet the fields
are massless: $c_{\mathrm{UV}}=c_{\mathrm{free}}$, $a_{\mathrm{UV}}=a_{%
\mathrm{free}}$. Instead, the IR fixed points of these flows are trivial: $%
c_{\mathrm{IR}}=a_{\mathrm{IR}}=0$. The quantities $\Delta a$ and $\Delta
a^{\prime }$ are calculable exactly and $\Delta a^{\prime }$ happens to
coincide with $\Delta c$. (A study of this coincidence can be found in ref. 
\cite{inv}.) Then (\ref{ineq}) implies $c_{\mathrm{free}}=\Delta c=\Delta
a^{\prime }\geq \Delta a=a_{\mathrm{free}}$ and therefore (\ref{freeineq}).

This argument does not generalize to vector fields and differential forms,
because there exists no RG flow connecting the free vector with the empty
theory and having $\Delta a^{\prime }=\Delta c$. The Proca theory of massive
vectors is singular in the ultraviolet and has $\Delta a^{\prime }=\infty $,
while $\Delta c$ is finite. This allows $a_{\mathrm{vector}}$ to be greater
than $c_{\mathrm{vector}}$, avoiding any contraddiction with our predictions.

An argument can be given to explain, to some extent, why vectors should
better have $a>c$. Let us recall that there exists a remarkable subclass of
conformal field theories having $c=a$ \cite{cea}. The equality $c=a$ is all
but difficult to fulfil, even at the free-field level. Hovewer, if both
scalars, fermions and vectors (or $(n-1)$-differential forms) had $a<c$,
there would exist no free-field theory with $c=a$. So, having proved that
scalars and fermions behave one way, it follows that vectors should behave
the opposite way, if we want that the $c=a$ theories are not so rare.

\bigskip

The considerations of the previous sections apply to unitary flows. It is
easy to check that, indeed, the inequality $a\leq c$ is violated in
non-unitary Gaussian fields with a $\Box ^{2}$ kinetic term. In \cite{inv}
it was shown that the equality $\Delta a^{\prime }=\Delta c$ is satisfied in
the massive higher-derivative models, if the masses are chosen
appropriately. Using the results of \cite{fest} and \cite{inv} we can see
that a higher-derivative scalar field with a $\Box ^{2}$-kinetic term has 
\[
\begin{tabular}{|c|c|c|c|c|c|c|}
\hline
$d$ & 4 & 6 & 8 & 10 & 12 & 14 \\ 
$c$ & $-8$ & $-5$ & $-4$ & $-\frac{7}{2}$ & $-\frac{16}{5}$ & $-3$ \\ 
$a$ & $-\frac{28}{3}$ & $-\frac{16}{9}$ & $-\frac{52}{45}$ & $-\frac{124}{135%
}$ & $-\frac{56302}{70875}$ & $-\frac{30544}{42525}$%
\end{tabular}
\]
etc. The equality $a\leq c$ holds only in $d=4$.

\bigskip

In summary, the understanding of section 5 explains some facts that
otherwise would appear to have no reason, namely why free scalars and
fermions have $a\leq c$, while the Proca theory of vector fields is singular
and vectors can have $a>c$.

\section{Irreversibility of the RG flow in odd (and even) dimensions}

\setcounter{equation}{0}

I this section I formulate a more general notion of irreversibility of the
RG\ flow, which applies also to odd dimensions and does not require the
existence of an $a$ function. I give a set of axioms (``oriented-triangle
inequalities'') that imply irreversibility without a global $a$ function and
the existence of a local $a$ function. I emphasize the conceptual
differences between irreversibility in even-dimensional and odd-dimensional
theories and study other topological and metric properties of the spaces $%
\mathcal{C}$ and $\mathcal{F}$, $\pi \mathcal{C}$ and $\pi \mathcal{F}$.

\bigskip

\textbf{Length and distance in odd dimensions.} The quantity $\Delta
a^{\prime }$ is meaningful and non-negative in the most general quantum
field theory \cite{at6d} and so the definition (\ref{length}) of length $%
L(\digamma )$ of the flow $\digamma $ applies also in odd dimensions. The
factor in front of the integral of formula (\ref{length}) has no meaning in
odd dimensions. Using $\Delta a^{\prime }$, the distance between the fixed
points can still be defined as the minimum (or the inferior limit, if the
minimum does not exist) of $L(\digamma )$ in the space of flows $\digamma $
connecting the same fixed points. Therefore, (\ref{distance}) holds also in
odd dimensions. Furthermore, it is still reasonable to conjecture that the
geodesic flows are classically conformal, but not all of the classically
conformal flows are geodesic in odd dimensions (see next section).

Instead, in odd dimensions it is not possible to associate a central charge $%
a$ to the fixed points \cite{fest} and express the minimum of $\Delta
a^{\prime }$ as the difference between the values of $a$ at the fixed
points. So, (\ref{auto}) is not ensured to hold. A global $a$ function might
not exist. The positivity of the minimum of $\Delta a^{\prime }$, i.e. the
positivity of a distance, is an obvious statement and does not imply that
the RG flow is irreversible. As a consequence, the irreversibility of the RG
flow in odd dimensions cannot be formulated as in even dimensions.

\bigskip

\textbf{Oriented distance and oriented triangles.} I have shown in section 6
that in even dimensions the triangle equality (\ref{treq}) holds, because
there exists a global $a$ function. In odd dimensions, instead, the distance
satisfies genuine triangle inequalities. With reference to Fig. 1, we have 
\begin{equation}
d_{13}\leq d_{12}+d_{23},\qquad d_{12}\leq d_{13}+d_{23},\qquad d_{23}\leq
d_{13}+d_{12}.  \label{trineq}
\end{equation}
The inequalities of the form $d_{ij}\geq |d_{ik}-d_{jk}|$ do not add
information, because they are implied by (\ref{trineq}).

The inequalities (\ref{trineq}) follow from the definition of distance.
Since non-trivial flows have an orientation, conventionally taken as $%
\mathrm{UV}\rightarrow \mathrm{IR}$, the distance is oriented. This allows
us to postulate more restrictive inequalities. With ``oriented triangle'' I
mean a triangle whose sides are oriented, unless they are associated with
trivial flows (zero length), in which case their orientation is unspecified.
If the flows of the triangle T are oriented as in Fig. 1, the
``oriented-triangle inequalities'' are 
\begin{equation}
d_{13}\leq d_{12}+d_{23},\qquad d_{13}\geq d_{23},\qquad d_{13}\geq d_{12}.
\label{trineq2}
\end{equation}

I conjecture that the unitary, renormalizable RG\ flows in arbitrary
dimensions satisfy the oriented-triangle inequalities. Observe that (\ref
{trineq2}) imply (\ref{trineq}), but (\ref{trineq}) do not imply (\ref
{trineq2}).

More generally: consider a sequence of flows $\digamma _{i}:\mathrm{C}%
_{i-1}\rightarrow \mathrm{C}_{i}$, $i=1,\cdots ,n$, with concordant
orientations, connecting two fixed points C$_{0}$ and C$_{n}$, with $n-1$
intermediate fixed points C$_{1}$, $\ldots $, C$_{n-1}$; then the distance $%
d(\mathrm{C}_{i},\mathrm{C}_{i-1})$ between two consecutive intermediate
fixed points is always smaller than (or equal to) the distance between C$%
_{0} $ and C$_{n}$. It is easy to see that this more general statement is
implied by the oriented-triangle inequalities. For the proof, it is
sufficient to apply the oriented-triangle inequalities to the set of
triangles $\mathrm{C}_{i-1}$-$\mathrm{C}_{i}$-$\mathrm{C}_{n}$, $i=1,\cdots
,n-1$.

\bigskip

\textbf{Irreversibility without a height function.} The irreversibility of
the RG\ flow is the statement that given a unitary flow $\digamma $
connecting the UV conformal field theory $\mathrm{C}_{1}$ to the IR
conformal field theory $\mathrm{C}_{2}$ there exists no unitary flow having $%
\mathrm{C}_{2}$ as UV fixed point and $\mathrm{C}_{1}$ as IR fixed point. In
particular, this means that the reversed flow $\overline{\digamma }$ of $%
\digamma $, obtained by scale inversion, which is the transformation 
\[
\mu \rightarrow \frac{1}{\mu },\qquad m\rightarrow \frac{1}{m}, 
\]
where $\mu $ denotes the dynamical scale and $m$ a generic dimensionful
parameter, does not exist or violates basic principles of quantum field
theory (typically unitarity or locality).

Strictly speaking, there is no need of a ``height'' ($c$ or $a$) function to
have irreversibility. If there exists a height function, irreversibility
follows \textit{a fortiori}, but if there does not exist a height function,
the RG flow can still be irreversible in the more general sense just
explained. Using the more general definition, it is easier to disprove
irreversibility, eventually: it is sufficient to find two RG\ flows with
interchanged the UV\ and IR limits interpolating between the same pair of
fixed points. At present, no example of this kind is known in the
literature, to my knowledge.

If a global $a$ function does not exist (generically, there might still
exist a local $a$ function: see below) then it is not possible to define a
counter of the (massless) degrees of freedom of the theory. Moreover, the
projected spaces $\pi \mathcal{C}$ and $\pi \mathcal{F}$ are not necessarily
one-dimensional: there might exist non-degenerate triangles.

\bigskip

\textbf{Irreversibility from oriented-triangle inequalities.} I now prove
that the oriented-triangle inequalities imply the irreversibility of the RG\
flow without a height function. Suppose that the flow $\digamma _{13}$ of
the triangle T is trivial ($d_{13}=0$, see Fig. 2a). Then (\ref{trineq2})
imply that the triangle is trivial. Applying this property to the case C$%
_{1}=\mathrm{C}_{3}$, we conclude there cannot exist two RG\ flows
connecting interchanged fixed points, namely the situation depicted in Fig.
2b is inadmissible. This means precisely that the RG flow is irreversible
without a height function. Similarly, a triangle with the orientation shown
in Fig. 2c is incompatible with (\ref{trineq2}), unless the triangle is
trivial.

\begin{figure}[tbp]
\centerline{\epsfig{figure=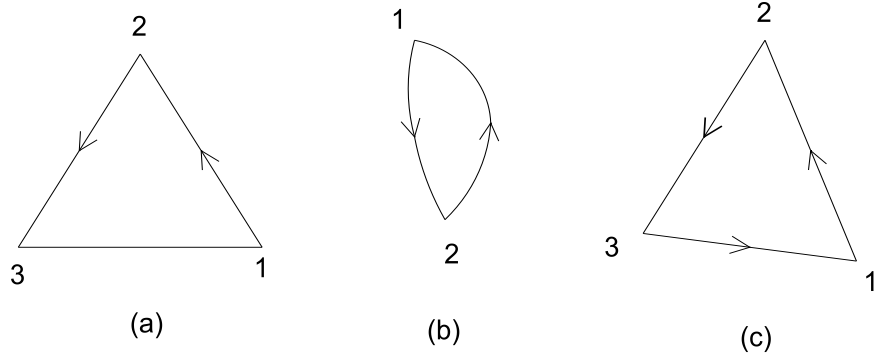,height=5cm,width=12cm}}
\caption{Inadmissible configurations}
\end{figure}

\bigskip

\textbf{Implication of a local }$a$\textbf{\ function from irreversibility
without an }$a$\textbf{\ function.} Here I prove that the oriented-triangle
inequalities imply also the existence of an $a$ function in the smooth
regions of the space $\mathcal{F}$.

I\ have explained in section 6 that the topology of the space $\mathcal{F}$
is expected to be involved. Let us consider, however, a two- (or higher-)
dimensional subset $\mathcal{R}$ of $\mathcal{C}$, and assume that for every
pair of points of $\mathcal{C}$ there exists a flow in $\mathcal{F}$
(eventually trivial, i.e. of zero lentgh), or a sequence of concordant
flows, connecting them. I want to show that under these smoothness
assumptions, the oriented-triangle inequalities (\ref{trineq}) imply the
existence of an $a$ function in $\mathcal{R}$.

Consider a point C in $\mathcal{R}$. Viewed from C, the set $\mathcal{R}$
can be written as the union of three subsets: 
\begin{equation}
\mathcal{R}=\mathcal{R}_{+}^{C}\cup \mathcal{R}_{0}^{C}\cup \mathcal{R}%
_{-}^{C}.  \label{decomp}
\end{equation}
$\mathcal{R}_{+}^{C}$ is the set of points P of $\mathcal{R}$ such that the
flows connecting P to C are oriented from P to C. $\mathcal{R}_{-}^{C}$ is
the set of points P of $\mathcal{R}$ such that the flows connecting P to C
are oriented from C to P. The set $\mathcal{R}_{0}^{C}$ is a surface of
points having zero distance from C.

The decomposition (\ref{decomp}) of $\mathcal{R}$ is well-defined because
the oriented-triangle inequalities imply irreversibility without an $a$
function, which means that given two points C$_{1}$ and C$_{2}$ in $\mathcal{%
R}$ all of the flows connecting them are either oriented from C$_{1}$ to C$%
_{2}$ or from C$_{2}$ to C$_{1}$, or the distance between C$_{1}$ and C$_{2}$
is zero.

Moreover, the decomposition (\ref{decomp}) is made with respect to a
reference point $\mathrm{C}$. Similar decompositions with respect to every
reference point C of $\mathcal{R}$ are given(obviously, C belongs to $%
\mathcal{R}_{0}^{C}$). Now, I\ have proved above that the situation of Fig.
2a is inadmissible. This ensures that the decomposition (\ref{decomp}) does
not depend on the reference point C in $\mathcal{R}_{0}^{C}$, i.e. $\mathcal{%
R}_{+}^{C}=\mathcal{R}_{+}^{C^{\prime }},$ $\mathcal{R}_{0}^{C}=\mathcal{R}%
_{0}^{C^{\prime }}$, $\mathcal{R}_{-}^{C}\cup \mathcal{R}_{-}^{C^{\prime }}$
for every C and C$^{\prime }$ in $\mathcal{R}_{0}^{C}$.

The decomposition (\ref{decomp}) can be used to define a ``height'' function
in $\mathcal{R}$. The $\mathcal{R}_{0}$ surfaces are the surfaces with equal
height. The space $\mathcal{R}_{+}$ is higher than $\mathcal{R}_{0}$, while $%
\mathcal{R}_{-}$ is lower than $\mathcal{R}_{0}$, which means that the $a$
function has larger values in the points of $\mathcal{R}_{+}^{C}$ than in $%
\mathcal{R}_{0}^{C}$ and a larger value in $\mathcal{R}_{0}^{C}$ than in $%
\mathcal{R}_{-}^{C}$, for every C in $\mathcal{R}$. Moreover, the values of $%
a$ should be positive. Apart from this, the values assigned to the function $%
a$ are arbitrary.

The construction of the $a$ function, however, might not extend to the
entire spaces $\mathcal{C}$ and $\mathcal{F}$. Some pairs of points C$_{1},$C%
$_{2}\in \mathcal{C}$ might not admit a flow or sequence of concordant flows
connecting them, and in this case there is no way to determine which point
is higher and which is lower. It might still be possible to assign some $a$%
-values to such points consistently with the $a$-values assigned in the
smooth subsets of $\mathcal{C}$, but it is doubtful that this is more than
an academic exercise and it would be quite arbitrary to interpret the
resulting $a$ function as the ``counter of the massless degrees of freedom
of the theory''.

\medskip

In conclusion, the interpretations of $\Delta a^{\prime }$ as length of the
RG flow and its minimum as distance between the fixed points give the
triangle equalities (\ref{trineq}). Secondly, the idea that the distance is
oriented suggests to conjecture the more restrictive oriented-triangle
inequalities (\ref{trineq2}). Finally, the irreversibility of the RG flow
does not require the existence of an $a$ function and can be defined as the
statement that two RG flows connecting interchanged UV\ and IR\ fixed points
are trivial. The oriented-triangle inequalities imply irreversibility
without a global $a$ function in $\mathcal{F}$. In the next section I
perform a check of the oriented-triangle inequalities in three dimensions
and prove the existence of non-degenerate triangles in odd dimensions. The
results support the idea that the RG flow is irreversible in odd dimensions
without a global $a$ function.

\bigskip

\textbf{Marginality.} I point out that the triangle inequalities (\ref
{trineq}), and \textit{a fortiori} the oriented-triangle inequalities (\ref
{trineq2}), imply that the distance $d(C_{\mathrm{UV}},C_{\mathrm{IR}})$ is
invariant under exactly marginal deformations of the fixed points $C_{%
\mathrm{UV}}$ and $C_{\mathrm{IR}}$. In other words, $d(C_{\mathrm{UV}},C_{%
\mathrm{IR}})=d(C_{\mathrm{UV}}^{\prime },C_{\mathrm{IR}}^{\prime })$ if $C_{%
\mathrm{UV}}$ and $C_{\mathrm{IR}}$ are continuously connected to $C_{%
\mathrm{UV}}$ and $C_{\mathrm{IR}}$, respectively. More generally, if we
consider a triangle T with $d_{23}=0$, the inequalities (\ref{trineq2}) give 
$d_{13}\leq d_{12}$, $d_{13}\geq d_{12}$ and therefore $d_{12}=d_{13}$. In
this more general formulation of the statement, the conformal theories C$%
_{2} $ and C$_{3}$ do not even need to be continuously connected. Observe
that only the distance, i.e. $\min_{\digamma \in \mathcal{F}}\Delta
a^{\prime }(\digamma )$, is expected to be marginal, but $\Delta a^{\prime }$
is not (for other details, see sect. 3.2 of \cite{fest}).

Proceeding as in the end of section 6, marginality implies that the
``surfaces of equal height'' are disconnected from one another.

\section{A calculation in three dimensions}

\setcounter{equation}{0}

In ref.s \cite{largeN,largeN2} some classes of three-dimensional classically
conformal theories have been constructed. They provide a valid laboratory to
test ideas about strongly-coupled quantum field theory and irreversibility.
Those RG flows interpolate between the UV fixed points of four-fermion
models and are exactly integrable in the running couplings at each order of
the large $N$ expansion. In this section I test the conjecture (\ref{trineq2}%
) in these three-dimensional flows.

The four-fermion model is a power-counting non-renormalizable theory in
three dimensions. It can be renormalized using a construction due to Parisi 
\cite{parisi} (see also \cite{rosen}) in the large $N$ expansion, where $N$
is the number of fermions. Only one diagram, precisely the fermion bubble,
contributes to the leading order. The fermion bubble gives a finite
contribution that improves the large-momentum behaviors of propagators in
such a way that, after resumming the fermion bubbles, the subleading
corrections are power-counting renormalizable. The theories of ref.s \cite
{largeN,largeN2} are renormalized in a similar way.

\bigskip

\textbf{The models.} I focus on the fermion models with lagrangian 
\begin{equation}
\mathcal{L}=\sum_{i=1}^{N}\overline{\psi }_{i}\partial \!\!\!\slash\psi
_{i}+\sum_{j=1}^{rN}\overline{\chi }_{j}\partial \!\!\!\slash\chi
_{j}+\lambda \sigma \left( \sum_{i=1}^{N}\overline{\psi }_{i}\psi
_{i}+g\sum_{j=1}^{rN}\overline{\chi }_{j}\chi _{j}\right) .  \label{lala}
\end{equation}
The effective $\sigma $-propagator is generated by the fermion bubble in the
large $N$ limit. The coupling $\lambda $ is inert and $\lambda ^{2}N$ is
kept fixed in the $1/N$ expansion, while the coupling $g$ runs. The RG\ flow
is integrable in $g$ at each order of the $1/N$ expansion \cite{largeN2}.
The model is chiral invariant, namely invariant under 
\[
\psi \rightarrow \gamma _{5}\psi ,\qquad \chi \rightarrow \gamma _{5}\chi
,\qquad \sigma \rightarrow -\sigma , 
\]
and has a strong-weak coupling duality 
\[
r\rightarrow \frac{1}{r},\qquad g\rightarrow 1/g,\qquad \lambda \rightarrow 
\frac{\lambda }{\sqrt{r}},\qquad \psi \leftrightarrow \chi ,\qquad \sigma
\rightarrow \sigma g\sqrt{r}, 
\]
which is exact at each order of the $1/N$ expansion. With four-component
(Dirac) fermions we can take $0\leq g\leq 1$. The beta function reads 
\[
\beta _{g}={\frac{8}{3\pi ^{2}N}}{\frac{g(g^{2}-1)}{1+rg^{2}},} 
\]
while $\beta _{\lambda }=-\varepsilon \lambda /2$ ($\lambda _{\mathrm{B}%
}=\lambda \mu ^{\varepsilon /2}$). The fixed points are 
\[
\mathrm{UV}\ (g=0):\qquad \Sigma _{N}\otimes \Psi _{rN}\ ;\qquad \qquad
\qquad \mathrm{IR\ }(g=1):\qquad \Sigma _{N(1+r)}\ ; 
\]
where $\Psi _{N}$ denotes $N$ free fermions and $\Sigma _{N}$ is the
conformal field theory defined by the lagrangian 
\[
\mathcal{L}=\sum_{i=1}^{N}\overline{\psi }_{i}\left( \partial \!\!\!\slash%
+\lambda \sigma \right) \psi _{i}. 
\]

The flow is classically conformal, so we expect that the distance between
the UV and IR fixed points is equal to the length of the flow, defined as 
\[
L(r)=\frac{16\pi }{3}\int \mathrm{d}^{3}x|x|^{3}\langle \Theta (x)\,\Theta
(0)\rangle . 
\]
First I derive the trace-anomaly formula, then compute $\Delta a^{\prime
}(r)=L(r)$ and use this result to check (\ref{trineq2}).

\bigskip

\textbf{Regularization and renormalization. }The dimensional regularization
has to be modified adding an evanescent, RG invariant non-local term to the
lagrangian (\ref{lala}) \cite{largeN}, to avoid the appearance of $\Gamma
[0] $s. This complicates the study of $\Theta $, since the embedding of a
non-local term in external gravity is quite involved. Here I use a more
practical regularization convention. The theory is still extended to $%
3-\varepsilon $ dimensions, but the cut-off term 
\begin{equation}
\frac{1+rg^{2}}{2\Lambda }\left( \partial _{\mu }\sigma \right) ^{2}
\label{eva}
\end{equation}
is added to the lagrangian. The limit $\Lambda \rightarrow \infty $ is
performed after the $\varepsilon \rightarrow 0$ limit. The $\varepsilon
\rightarrow 0$ limit renormalizes the fermion loops, while the $\Lambda
\rightarrow \infty $ limit gives sense to the loops containing $\sigma $%
-propagators. The factor in (\ref{eva}) is chosen to have manifest duality
invariance.

I use a classically conformal minimal subtraction scheme. The poles and the
terms proportional to powers of $\ln \Lambda /\mu $ are subtracted away with
no finite part, as well as the linear, quadratic and cubic divergences ($%
\Lambda ^{k}$, $\Lambda ^{k}/\varepsilon ^{m}$, $\Lambda ^{k}(\ln \Lambda
/\mu )^{m}$ for $k>0$, $m\geq 0$). The renormalized lagrangian contains also
a counterterm of the form 
\begin{equation}
\Lambda \frac{1+rg^{2}}{2}\delta Z_{m}\sigma ^{2}  \label{mas}
\end{equation}
that cures the linear divergences. Chiral invariance forbids an analogous
term for the fermions. To avoid a heavy notation, I go through the
derivation as if (\ref{mas}) were not there. It is easy to include this term
and check that the result (\ref{teta}) is unmodified.

\bigskip

\textbf{The trace-anomaly operator equation.} I start from the general
integrated formula \cite{234} 
\begin{equation}
\left\langle \int \mathrm{d}^{3-\varepsilon }x\ \widehat{\Theta }%
(x)\right\rangle =-\mu \frac{\partial \Gamma }{\partial \mu }.  \label{inte}
\end{equation}
Here $\Gamma $ is the quantum action and $\widehat{\Theta }$ is the trace of
the stress tensor up to terms proportional to the field equations (which are
irrelevant for the computation of $\Delta a^{\prime }$). Formula (\ref{inte}%
) says that the insertion of an integrated trace is equal to an insertion of 
$-\mu S\partial /\partial \mu $, where $S$ is the action. Since $\mu \mathrm{%
d}\Gamma /\mathrm{d}\mu =0$, we can use the Callan-Symanzik equations and
rewrite $-\mu \partial /\partial \mu $ as 
\begin{equation}
\beta _{\lambda }\frac{\partial }{\partial \lambda }+\beta _{g}\frac{%
\partial }{\partial g}.  \label{CS}
\end{equation}
Now, $\beta _{\lambda }$ is evanescent and $\partial /\partial \lambda $ is
a renormalized operator, because the derivative $\partial /\partial \lambda $
of a renormalized correlation function is obviously finite. Therefore the
piece $\beta _{\lambda }\partial /\partial \lambda $ can be omitted. In
summary, the $\ln \mu $-derivative of a correlator can be re-expressed as
minus its $g$-derivative times $\beta _{g}$.

The differentiation of a correlation function with respect to $g$ is
equivalent to the insertion of the integrated operator $-\partial S/\partial
g$. The $g$-derivative of the action is done keeping the renormalized
couplings and fields constant. Alternatively, we can keep the bare fields
constant, since the difference amounts to terms proportional to the field
equations. However, we do have to differentiate the bare parameters. Since $%
\lambda _{\mathrm{B}}$ does not depend on $g$, we just have to differentiate 
$g_{\mathrm{B}}$. The result is 
\[
\int \mathrm{d}^{3}x\ \widehat{\Theta }(x)=\beta _{g}\frac{\partial g_{%
\mathrm{B}}}{\partial g}\frac{\partial S}{\partial g_{\mathrm{B}}}=\beta _{g}%
\frac{\partial g_{\mathrm{B}}}{\partial g}\int \mathrm{d}^{3}x\ \left[
\lambda _{\mathrm{B}}\sigma _{\mathrm{B}}\sum_{j=1}^{rN}\overline{\chi }_{_{%
\mathrm{B}}j}\chi _{_{\mathrm{B}}j}+\frac{rg_{\mathrm{B}}}{\Lambda }\left(
\partial _{\mu }\sigma _{\mathrm{B}}\right) ^{2}\right] . 
\]
Using the $\sigma $ field equation, we obtain also 
\[
\frac{1}{\beta _{g}}\int \mathrm{d}^{3}x\ \widehat{\Theta }(x)={\frac{%
\lambda _{\mathrm{B}}}{1+rg_{\mathrm{B}}^{2}}}\frac{\partial g_{\mathrm{B}}}{%
\partial g}\int \mathrm{d}^{3}x\ \sigma _{\mathrm{B}}\left( \sum_{j=1}^{rN}%
\overline{\chi }_{_{\mathrm{B}}j}\chi _{_{\mathrm{B}}j}-rg_{\mathrm{B}%
}\sum_{i=1}^{N}\overline{\psi }_{\mathrm{B}i}\psi _{\mathrm{B}i}\right) . 
\]
Now, since the left-hand side is a renormalized operator, the insertion of
the right-hand side in a correlation function is finite. Therefore, the
right-hand since is the renormalized version of the operator obtained
suppressing the $_{\mathrm{B}}$s everywhere.

Finally, assuming that the integral can be taken away, the trace-anomaly
operator formula 
\begin{equation}
\widehat{\Theta }={\frac{\lambda \beta _{g}}{1+rg^{2}}}\sigma \left(
\sum_{j=1}^{rN}\overline{\chi }_{j}\chi _{j}-rg\sum_{i=1}^{N}\overline{\psi }%
_{i}\psi _{i}\right)  \label{teta}
\end{equation}
is obtained, where the right-hand side is understood to be the renormalized
composite operator. It is immediate to check that this expression is duality
invariant. (In \cite{largeN} a non-manifestly duality invariant expression
was given. The formula of \cite{largeN} differs from (\ref{teta}) by a term
proportional to the $\sigma $ field equation and an evanescent term coming
from (\ref{eva}), but gives exactly the same $\Delta a^{\prime }$.)

We have to justify that the integral can be freely taken away. Since we are
using a local regularization, $\widehat{\Theta }$ is local. $\widehat{\Theta 
}$ might differ from (\ref{teta}) by total derivatives. These can only be $%
\partial _{\mu }\left( \overline{\psi }\gamma _{\mu }\psi \right) $, $%
\partial _{\mu }\left( \overline{\chi }\gamma _{\mu }\chi \right) $ and $%
\Box \sigma ^{2}/\Lambda $. The first two terms are absent. This can be seen
observing that $\Theta $ is invariant under charge conjugation, but $%
\partial _{\mu }\left( \overline{\psi }\gamma _{\mu }\psi \right) $ and $%
\partial _{\mu }\left( \overline{\chi }\gamma _{\mu }\chi \right) $ are not.
The term $\Box \sigma ^{2}/\Lambda $ is a renormalized evanescent total
derivative, so it gives no contribution when the cut-off is removed.

$\bigskip $

$\Theta $\textbf{\ two-point function and length of the flow.} Using the
techniques of \cite{largeN2}, we know that in the leading-log approximation
it is sufficient to compute the two-point function of the operator $\sigma
\left( \overline{\chi }\chi -rg\overline{\psi }\psi \right) $ to the leading
order. The relevant diagrams are depicted in Fig. 3. The sum of diagrams (b)
and (c) vanishes. It remains to compute the diagram (a), which is
straightforward in the $x$ space. We obtain 
\[
\langle \Theta (x)\,\Theta (0)\rangle ={\frac{%
32rg^{2}(1/|x|)(1-g^{2}(1/|x|))^{2}}{9\pi ^{8}N^{2}x^{6}(1+rg^{2}(1/|x|))^{4}%
}}, 
\]
where $g(1/|x|)$ is the running coupling. From this expression it is
immediate to derive $\Delta a^{\prime }$. The result is 
\[
\Delta a^{\prime }(r)=\frac{16\pi }{3}\int \mathrm{d}^{3}x|x|^{3}\langle
\Theta (x)\,\Theta (0)\rangle ={\frac{64}{9\pi ^{4}N}}\left( 1-{\frac{1}{r+1}%
}\right) \equiv {\frac{64}{9\pi ^{4}N}}d(r). 
\]
The distance $d(r)$ between the fixed points has been defined eliminating an
irrelevant factor. In the limit $r\rightarrow 0$ the order $\mathcal{O}(1/N)$
of $\Delta a^{\prime }(r)$ tends to zero. However, this does not mean that
the flow is trivial. Indeed, it was proved in \cite{largeN2} that the limit $%
r\rightarrow 0$ exists and is a non-trivial RG flow. Therefore we expect
that the subleading orders give $\Delta a^{\prime }>0$ also in this case. 
\begin{figure}[tbp]
\centerline{\epsfig{figure=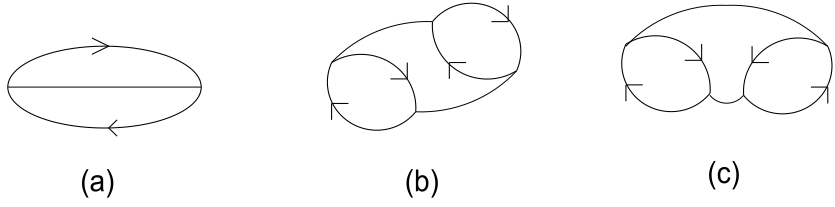,height=3cm,width=11cm}}
\caption{Diagrams for the calculation of $\Delta a^{\prime }$}
\end{figure}

\bigskip

\textbf{Test of the prediction.} Consider the theory

\begin{equation}
\mathcal{L}=\sum_{i=1}^{N}\overline{\psi }_{i}\partial \!\!\!\slash\psi
_{i}+\sum_{j=1}^{r_{1}N}\overline{\chi }_{j}\partial \!\!\!\slash\chi
_{j}+\sum_{k=1}^{r_{2}N}\overline{\zeta }_{k}\partial \!\!\!\slash\zeta
_{k}+\lambda \sigma \left( \sum_{i=1}^{N}\overline{\psi }_{i}\psi
_{i}+g_{1}\sum_{j=1}^{r_{1}N}\overline{\chi }_{j}\chi
_{j}+g_{2}\sum_{k=1}^{r_{2}N}\overline{\zeta }_{k}\zeta _{k}\right) .
\label{theory}
\end{equation}
Three sets of fermions are coupled together by means of two couplings $g_{1}$
and $g_{2}$, with $0\leq g_{1,2}\leq 1$. The fixed points are $\left(
g_{1},g_{2}\right) =(0,0)$, $(0,1)$, $(1,0)$, $(1,1)$. We study the triangle
of flows 
\begin{eqnarray}
&&\digamma _{12}:\quad g_{2}\equiv 0,\quad 0\leq g_{1}\leq 1\ ;\qquad \qquad
\digamma _{13}:\quad g_{1}\equiv g_{2},\quad 0\leq g_{1}\leq 1\ ;\qquad 
\nonumber \\
&&\qquad \qquad \qquad \qquad \qquad \digamma _{23}:\quad g_{1}\equiv
1,\quad 0\leq g_{2}\leq 1\ .  \label{values}
\end{eqnarray}
with conformal fixed points 
\[
\mathrm{C}_{1}=\Sigma _{N}\otimes \Psi _{N(r_{1}+r_{2})}\ ;\qquad \mathrm{C}%
_{2}=\Sigma _{N(1+r_{1})}\otimes \Psi _{Nr_{2}}\ ;\qquad \mathrm{C}%
_{3}=\Sigma _{N(1+r_{1}+r_{2})}\ . 
\]

The distances 
\[
d_{12}=1-{\frac{1}{1+r_{1}}},\qquad d_{13}=1-{\frac{1}{1+r_{1}+r_{2}}}%
,\qquad d_{23}=1-{\frac{1+r_{1}}{1+r_{1}+r_{2}}}, 
\]
do satisfy the oriented-triangle inequalities (\ref{trineq2}). This is an
indication in favor of irreversibility in odd dimensions, in non-trivial
agreement with the understanding offered in this paper.

The results imply also that there exist non-degenerate triangles in odd
dimensions, and therefore it is impossible to define a global $a$ function
such that $d(\mathrm{C}_{\mathrm{UV}},\mathrm{C}_{\mathrm{IR}})=a(\mathrm{C}%
_{\mathrm{UV}})-a(\mathrm{C}_{\mathrm{IR}})$.

Another consequence is that not all classically conformal flows are
geodesic. Indeed, the theory (\ref{theory}) depends on two parameters.
Choosing more generic values of the couplings than the particular cases (\ref
{values}) it is possible to fill the triangle with flows that interpolate
continuously between $\digamma _{13}$ and $\digamma _{12}$+$\digamma _{23}$.
All of the flows are classically conformal, but their lengths vary
continuously from $d_{13}$ to $d_{12}+d_{23}>d_{13}$.

Conversely, it is still plausible that all geodesic flows are classically
conformal.

\section{Conclusions}

In this paper I\ have studied several aspects of the irreversibility of the
RG flow, and elaborated a conceptual picture that is consistent with the
present knowledge and explains some facts that othewise would appear to be
somewhat mysterious. The investigation has direct connections with different
research domains, such as the study of the topological and metric properties
of the spaces of RG\ flows and conformal field theories. Little is known
today about these difficult subjects. It is often compulsory to proceed
empirically, or by means of conjectures and cross-checks, or axioms and
logical implications. On the other hand, it is obvious that before spending
a lot of effort to obtain rigorous proofs, it is better to have a clear idea
of what might be worth trying to prove. I am convinced that the results of
this paper are a good starting point to address the future research in this
area.

In the most general terms, irreversibility is the statement that there exist
no pairs of non-trivial flows of unitary theories connecting interchanged
UV\ and IR fixed points. A primary goal of the paper was to elaborate a
simple and clear set of axioms that imply irreversibility. It is worth to\
recapitulate the guidelights of the arguments. The scheme-invariant area $%
\Delta a^{\prime }$ of the graph of the effective beta function between the
fixed points is taken as definition of the length of the RG flow. Then the
minimum of $\Delta a^{\prime }$ in the space of flows connecting the same UV
and IR fixed points defines the distance between the fixed points. Since the
flows are oriented, the distance is oriented and it is possible to postulate
``oriented-triangle'' inequalities.

These notions form the ``axioms of irreversibility'', because they imply the
irreversibility of the RG flow, in even and odd dimensions. At the moment, I
do not have a proof that (unitary) quantum field theory does satisfy the
axioms of irreversibility, but I\ can test some non-trivial consequences of
those axioms. In section 9 I\ have studied certain triangles made of
classically-conformal flows in three dimensions and showed that the
oriented-triangle inequalities are fulfilled. At the same time, those
results show that there exist non-degenerate triangles in odd dimensions.

In even dimensions, more powerful tools are available, in particular there
exists a candidate global $a$ function for irreversibility. This and other
arguments lead to a further conjecture, namely that in even dimensions the
oriented distance between the fixed points coincides precisely with $\Delta
a $. Again, a definitive proof of this conjecture is not available, but
checks of its consequences are possible. In particular, the chain of
inequalities $\Delta a^{\prime }\geq \Delta a\geq 0$ and a few other facts
imply that free massive scalars and fermions (but not vectors) always have $%
c\geq a$, which is true.

I recall that the existence of a global $a$ function with the mentioned
properties implies that all triangles are degenerate in even dimensions.

The no-go statements of section 2 are of course rigorously proved, but there
it was easy to give proofs, since it was sufficient to exhibit
couter-examples.

\bigskip

As said, one of the purposes of the investigation of this paper is to make a
first attempt to characterize the space of conformal field theories and
flows in higher dimensions. Several subspaces, i.e. classes of flows and
conformal theories, need to be classified and characterized. In the long
range, the final goal of this kind of investigation is to establish with
sufficient precision to which classes QCD, the Standard Model and Gravity
belong, and explain their phenomenological properties, maybe also
quantitatively, using this information.

The difficulties of this kind of research smear out when theories in various
dimensions are compared with one another. In particular, it is useful to
compare even and odd dimensions and, in even dimensions, dimension two and
dimension greater than two. Important tools for the classification of fixed
points and flows are the definitions of length of the RG flow, distance
between the fixed points, oriented distance and irreversibility, with and
without a global $a$ function. In odd dimensions a global $a$ function does
not exist, but a global $a$ function is not necessary to have
irreversibility. I believe that irreversibility holds also in odd
dimensions, in the more general sense elaborated here.

The irreversibility of the RG flow has a variety of implications. For
example, in even dimensions, where the counter $a$ of degrees of freedom is
globally defined, irreversibility might explain why quantizing the theories
from the IR is often problematic: it is reasonable to expect that climbing
against the stream of irreversibility, the missing degrees of freedoms
should be added manually. This might be the reason why $i$) QED has the
Landau pole; $ii$) the $\varphi _{4}^{4}$ theory is probably trivial; $iii$)
gravity -- seen from the IR -- is non-renormalizable; $iv$) examples of
IR-free, UV-interacting conformal windows in even dimensions are not known.
Instead, in odd dimensions there exist IR-free, UV-interacting RG flows and
their quantization does not exhibit particular difficulties, maybe because
it is possible to interpolate between the fixed points exactly in the
running couplings at each order of the large N expansion \cite{largeN2}.

The results of this paper stress once again that to fully understand the
properties of quantum field theory we need more powerful tools than the ones
we are accostumed to, a new framework and maybe a new language. Hopefully,
quantum field theory is going to please us with some interesting surprises
in the future.

\vskip .6truecm \textbf{Acknowledgements}

\vskip .3truecm

I thank the Aspen Center for Physics for warm hospitality during the early
stage of this work and J. Evslin, M. Mintchev, P. Menotti and F. Landolfi
for discussions.

\end{document}